\documentclass[12pt]{article}
\usepackage{amssymb,amsmath,graphicx}
\usepackage{epsf}
\usepackage{Iopconf}
\usepackage{bm}
\usepackage{amsfonts}
\begin{document}
\def\[{\bigl[}
\def\]{\bigr]}
\def\({\bigl(}
\def\){\bigr)}
\def\p{\partial}
\def\o{\over}
\def\cm{\cal M}
\def\R{\bf R}
\def\be{\begin{equation}}
\def\ee{\end{equation}}
\def\bea{\begin{eqnarray}}
\def\eea{\end{eqnarray}}
\def\nn{\nonumber}
\def\pd{\partial}
\def\a{\alpha}
\def\b{\beta}
\def\g{\gamma}
\def\d{\delta}
\def\m{\mu}
\def\n{\nu}
\def\t{\tau}
\def\l{\lambda}
\def\L{\Lambda}
\def\s{\sigma}
\def\e{\epsilon}
\def\scri{\mathcal{J}}
\def\cM{\mathcal{M}}
\def\tcM{\tilde{\mathcal{M}}}
\def\RR{\mathbb{R}}
\def\CC{\mathbb{C}}

\begin{flushright}
IFT-UAM/CSIC-04-55\\ hep-th/0410288\\
\end{flushright}

\title
{The supermembrane with central charge \\ 
as a bundle of D2-D0 branes}
\author
  {M. P. Garc\'\i a del
Moral\dag\footnote{E-mail: {\tt
gdelmoral@delta.ft.uam.es}}, A. Restuccia\ddag\footnote{E-mail:
 {\tt arestu@usb.ve}}}
\affil{\dag\ Departamento de F\'{\i}sica Te\'orica, C-XI,
 Universidad Aut\'onoma de Madrid,
 E-28049-Madrid, Spain}
\affil{\ddag\ Departamento de F\'\i sica,
 Universidad Sim\'on Bol\'ivar, Caracas, Venezuela}
\beginabstract
We discuss the consistency of the $D=11$ supermembranes with non zero
central charge arising from a nontrivial winding CSNW. The spectrum of
its regularized Hamiltonian is discrete ans its heat kernel in terms of
a Feynman formula may be rigorously constructed. The $N\to\infty$ limit is discussed. Since CSNW is equivalent to a
noncommutative supersymmetric gauge theory on a general Riemann surface,
its consistency provides a proof that all of them are well defined quantum theories. We
interpret the supermembrane with central charge $n$, in the type
$IIA$ picture, as a bundle of D2 branes with $n$ units of D0 charge induced by a nonconstant magnetic flux.
\endabstract
\section{Introduction}
The nonperturbative contributions to the superstring theory arise from the quantization of
M-theory in 11 dimensions. Related to it, is the problem of quantization of 11 dimensional
supermembranes. Recent progress to this problem has been done in the context of the quantization
 of supermembranes with nontrivial central charge \cite{MOR,MR,M1,M2,M3,JR,BR}.
 This is a well defined sector of the complete theory. This sector is also related,
 as we discuss in the paper, with noncommutative gauge theories over Riemann surfaces
 of genus $g\ge 1$. The quantum consistency of supermembranes with fixed central charges
 would provide an indirect proof of consistency of all these noncommutative gauge theories.

In order to realize the nontrivial central charge condition, there
must be a minimal inmersion from $\Sigma$ to the target space
which implies that, for the case of flat target spaces, the
worldvolume of the supermembrane is a calibrated submanifold of
the target space. Calibrations, and its generalizations, have been
extensively used to describe supersymmetric branes in supergravity
backgrounds \cite{GCU}. Minimal inmersions can
also describe non-BPS minimal solutions \cite{JR}.

Supermembranes are extended objects of $2+1$ dimensions, which
live in 11D. They were originally proposed over a $D=11$ Minkowski
target space as candidates to fundamental objects in the context
of M-theory. The continuity of the spectrum \cite{DWLN} lead to reinterpret
them as a many body theory. This property relies on two basic
facts: first, the presence of singular configurations with zero
energy at classical level, and second, supersymmetry. Under
compactification this behaviour seems to remains the same \cite{DWPP},
although as far as we know a rigurous proof has not yet been
done.\\ 
In the following we will summarize the main properties of
the supermembrane with central charges. This theory exhibits a
completely different behaviour from the previous one: it does not
contain string-like spikes, its supersymmetric spectrum at quantum
level is purely discrete, and its heat kernel is perfectly well
defined in terms of a Feynman formula in contrast to the supermembrane
on a Minkowski target space.\\ 
The supermembrane with
central charge due to the winding is equivalent to a symplectic
noncommutative Super Yang Mills theory over a Riemanian surface of
genus $g\ge 1$. The theory is invariant under area preserving
diffeomorphisms, which in 2 dimensions coincide with the
symplectomorphisms. The symplectic 2-form arising from the central
charges, for $g\ge 2$ cannot be globally expressed in terms of a
constant 2-form (as in the case of $g=1$). However, by using
Darboux theorem, we can reduce the symplectic two-form to a
constant antisymmetric tensor on each open set of a Darboux covering of
$\Sigma$ and hence to have a conventional noncommutative gauge
theory at each open set of the covering, all patched together by
symplectomorphisms.\\
In the dual picture we will discuss its
interpretation in terms of D2 branes with fluxes. Fluxes are
induced by the nontrivial winding and not by turning on a
constant 3-form which is reduced in ten dimensions to a constant
B field. These fluxes charge the D2 branes with $n$ units of D0 monopole
charge, appearing a system of D2-D0 branes over the manifold.
This mechanism imposes constraints on the
target dimension where this system of D2-D0 branes can appear.
It is a well known fact that supersymmetry preserving a set of
intersecting branes may merge to form a single brane on a smooth
calibrated surface \cite{smith, helling}. It allows to interpret the
supermembrane with central charge in the dual picture as intersecting
branes which recombine over the target manifold in this bundle of D2-D0
branes. The scalars parametrizing the position of this bundle
in the transverse space becomes massive through the recombination
process as a kind of Higgs mechanism. The effect is a bundle of D2 brane
charged over the compactification manifold.\\ 
In the case where there are
several M2 branes CSNW minimally inmersed in different 2-cycles
contained in an enough rich target manifold (with singularities,
cycles) we speculate that it can lead to phenomenological models. In
this interpretation each of these objects is thought as a
fundamental object over a particular nontrivial 2-cycle and
propagating in a 4D Minkowski space-time. The interpretation of the
wrapped supermembrane as a fundamental object was also
anticipated by Kallosh \cite{kallosh}, based on a semiclassical
quantization of the supermembrane done in \cite{stelle}. This semiclassical analysis in the
light of our results of the quantum spectrum recover sense as a limit.
\section{Supermembrane with non-trivial central
charge}
We consider the $D=11$ supermembrane in the light cone
gauge $\cite{DWHN}$. In this gauge the potential is given by
\be
V(X)
=\{X^{M}, X^{N}\}^{2}\quad M,N=1,..,9
\ee
where
\be
\{X^{M}, X^{N}\}^{2}= \frac{\epsilon ^{ab}}{\sqrt{W(\sigma)}}\partial_{a}X^{M}\partial_{b}X^{N}.
\ee
The scalar density $\sqrt{W(\sigma)}$ is introduced in the formulation by the partial gauge fixing procedure \cite{DWHN} and will take it to be the volume of the minimal inmersion which is going to be introduced shortly.
$X^{M},M=1,..,9$ are maps from a Riemann surface $\Sigma$, a torus in the discussions of this setting, to the target space which we assume to be $M_{7}\times S^{1}\times S^{1}$, $M_{7}$ being a seven dimensional Minkowski space-time.\\
The generalization to more general target spaces and Riemannian surfaces has been considered in \cite{JR}. We take $X_{r},r=1,2$ to be maps from $\Sigma$ to $S^{1}\times S^{1}$ and $X^{m},m=1,..,7$ the maps from $\Sigma$ to $M_{7}$.\\
The maps $X_{r}$ satisfy the conditions
\be
\oint _{c_{i}}dX_{r}=2\pi m_{ri}\quad r=1,2 \ee
where $C_{i}$ is a
basis of homology over $\Sigma$. These winding conditions ensure
that each $X_{r}, r=1,2$ is a map from $\Sigma \to S^{1}$. In
order that the image of $\Sigma$ by $X_{r}, r=1,2$ describe
also a torus, one has to impose
\be
Z=\int_{\Sigma}(dX_{r}\wedge dX_{s})\epsilon^{rs}= 2\pi n\neq 0
\ee
where $n=det (m_{jr})$ and $r=1,2$ and also $j=1,2$ since $\Sigma$
is also a torus. We notice that the condition $Z=2\pi n \neq 0$
corresponds to have a nontrivial central charge on the
supersymmetric algebra of the supermembrane. Between all the maps
from the torus $\Sigma$ to the target space satisfying (4) there
is one which minimizes the hamiltonian of the supermembrane. It is
realized in terms of the basis of harmonic one-forms over the
torus $\Sigma$, which we denote $dX^{r}, r=1,2$. \\ Any one-form 
over $\Sigma$ may be rewritten as
\be
dX^{r}=m^{r}_{s}d\widehat{X}^{s}+\delta^{rs}dA_{s} \ee
where
$A_{s}$ is a single-valued object over $\Sigma$ and $\det
m^{r}_{s}=1$. Using the residual invariance, the area preserving
diffeomorphisms not connected with the identity we may fix at
$m^{r}_{s}=\delta^{r}_{s}$. The transverse coordinates to the
supermembrane $X^{m},m=1,..,7$ we assumed to be single-valued
over $\Sigma$ since they are valued on Minkowski D=7.\\
The hamiltonian of the supermembrane with nontrivial central
charge may now be rewritten in terms of $X^{m}, m=1,..,7$ and
$A_{r}, r=1,2$. The resulting expression is:
\begin{equation}\label{e5} 
 \begin{aligned} 
H&=\int_{\Sigma}\frac{1}{2}\sqrt{W}[P^{2}_{m}+ \Pi^{2}_{r}+
\frac{1}{2}W\{X^{m},X^{n}\}^{2}+W(\mathcal{D}_{r}X^{m})^{2}+\frac{1}{2}W(\mathcal{F}_{rs})^{2}]
\\&+\int_{\Sigma}[\frac{1}{8}\sqrt{W}n^{2}-\Lambda
(\mathcal{D}_{r}\Pi_{r}+\{X^{m},P_{m}\})]\\& + \int_{\Sigma}
\sqrt{W} [- \overline{\Psi}\Gamma_{-} \Gamma_{r}
\mathcal{D}_{r}\Psi +\overline{\Psi}\Gamma_{-}
\Gamma_{m}\{X^{m},\Psi\}+
 \Lambda \{ \overline{\Psi}\Gamma_{-},\Psi\}]
 \end{aligned} 
 \end{equation} 
where $P_{m}$ and $\Pi_{r}$ are the conjugate momenta to $X^{m}$
and $A_{r}$ respectively. $\mathcal{D}_{r}$ and $\mathcal{F}_{rs}$
are the covariant derivative and curvature of a symplectic
noncommutative theory \cite{MOR,M2}, constructed from the
symplectic structure $\frac{\epsilon^{ab}}{\sqrt{W}}$ introduced by the central charge. The last term represents its
supersymmetric extension in terms of Majorana spinors. $\sqrt{W}$
is the worldvolume constructed from the minimal inmersion
$X^{r}, r=1,2$ of $\Sigma \to S^{1}\times S^{1}$. This is the
requirement \cite{JR}, in order to extend the construction to more
general target spaces. The image of $\Sigma$ under the minimal
inmersion is a calibrated submanifold.\\ In the general
situation where the genus of $\Sigma$ is $\ge 2$ the symplectic
two-form may only be considered constant, via Darboux theorem, in
an open neighbourhod $U$ over $\Sigma$ which cannot be extended
globally to all $\Sigma$. Only in the case when $\Sigma$ is a
torus the constant global extension is possible. The
noncommutative theory we obtain may be thought as a set of
noncommutative gauge theories generated
by a constant $B$ field, on each open set $U$ of
a Darboux covering of $\Sigma$, all patched together. There is no 
Seiberg-Witten  \cite{SW} limit involved. This is an  interesting point since
CSNW contain within its configuration space the superstrings with winding.
\section {Discretness of the spectrum and the Heat Kernel}
The hamiltonian (\ref{e5}) may be regularized \cite{M1} making use of the
property, we have already commented, that the harmonic modes may
be fixed by using the area preserving diffeomorphisms not
connected to the identity \cite{JR}. We are left with the area
preserving diffeomorphisms connected to the identity which are
generated by the Gauss constraint of the noncommutative theory.\\
Once the regularized hamiltonian has been obtained one may proceed
to analyse its spectrum. The bosonic potential $V_{B}$ has the
property that $V_{B}\to \infty$ as the point in the configuration
space goes to $\infty$. Furthermore $V_{B}=0$ implies $X=A=0$. By
$V_{B}$ we refer to the contribution of the noncommutative
Yang-Mills, without including the constant arising from the
contribution of the ground state. Those properties imply that the
spectrum of the bosonic hamiltonian is discrete \cite{M2},
moreover taking into account the structure of the fermionic
contribution it is possible to show that the fermionic potential
is a perturbation relatively bounded of $H_{B}$ \cite{BR}, it is
then compatible with it and the whole hamiltonian has also a
discrete spectrum \cite{M3}. The heat kernel for the
supersymmetric regularized hamiltonian $H$ has been constructed in
\cite{BR}. A Feymann-Kac formula has been obtained and its
convergence in terms of the strong operator topology has been
shown. An important result in \cite{BR} is related to the
constructions of the semigroup $\exp^{-tH}$ as an operator
belonging to the $C_{r}$ Neumann-Schatten class, a Banach space,
where $r \sim N$ is the integer associated to the $SU(N)$
truncation of the original theory. \\
Since the Banach spaces $C_{r}$ satisfy
\be
C_{1}\subset C_{2}\subset ... \subset C_{r}\subset C_{s}\subset
C_{\infty} \quad r<s \ee
it is expected for the supersymmetric
case that the $N\to\infty$ limit makes sense. The supersymmetry
should play a fundamental role in this limit in order to balance
the bosonic sector which manifestly $\exp^{-tH_{B}}\to 0$
when $N\to\infty$. This limit would extend rigurously our results
of the regularized hamiltonian of the supermembrane with central
charge to the original one. The limit of large N would define the
quantization of the supermembrane with central charge. We hope to
report on this limit soon.
 \section{The supermembrane with central charge as a bundle of D2-D0 branes}
 The supermembrane with nontrivial central charge exhibits the new
 properties that we have discussed in previous sections:
\par No singular configurations at classical level.
A noncommutative Super Yang-Mills theory defined at its worldvolume.
Discrete supersymmetric spectrum at quantum level.
A well defined heat kernel.
Massive transverse scalar fields that lift this moduli.
A BPS absolute  minimum.\\
All of these properties differ substantially from the case of the
$D=11$ supermembrane. There are some characteristics of the model
that could be emphasized to understand its properties. First, in
the compactification process we do not restrict in any way the
dynamics of the supermembrane, so for a given time, it remains
being a bidimensional surface for the whole target space. Second,
because of an appropiate handle of the multivalued forms
associated to the compactification process, we are able to extend
DWHN matrix model \cite{DWHN} to compactified spaces and obtain an
exact regularization of the hamiltonian in terms of $D0$ branes \cite{M1}.
All of the interaction terms have been taken into account and
emerge naturally from the hamiltonian.\\
In the dual picture, the
supermembrane with fixed central charge $n$ induced by the nontrivial
winding on a torus, can be interpreted in terms of a $D2$ brane
with $n$ units of nonconstant magnetic flux. This flux verifies the following properties,
\begin{equation}\label{e6} 
 \begin{aligned}
&F_{2}(\sigma)=\epsilon^{rs}dX^{r}(\sigma)\wedge dX^{s}(\sigma)\\
&dF_{2}=0\\
&\int_{\Sigma_{2}}d^{2}\sigma F_{2}=2\pi n
\end{aligned}
\end{equation}

\begin{figure}
\begin{center}
\centering
\epsfysize=1.2in
\leavevmode
\epsfbox{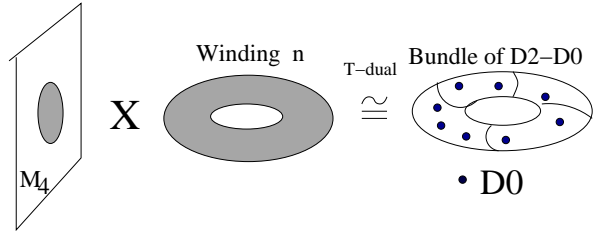}
\end{center}
\caption[]{\footnotesize Description of a supermembrane in Minkowski
 space with an irreducible winding $n$ on a Riemann surface of the
 transverse space. In the dual picture it is equivalent to a bundle of
 D2 branes with $n$ units of D0 branes attached and patched together 
around the compact manifold. }
\label{d2}
\end{figure}
The nontrivial winding is the quantization condition for the flux
induced by the dualized scalar fields $dX^{r}$ over the
worldvolume. The global condition represents a $2\pi n$ RR Dirac
monopoles. On a torus it can be interpreted as a system of
D2-D0 branes with $2\pi n$ units of $D0$ charge. We would like
to emphasize the physical origin of this flux in the context of
M-theory as purely topological. It is important to remark that since
this flux is not defined by any external background, this bundle of D2-D0
branes cannot exists in $10$ dimensions but in $9$ or less dimensions.\\
Formally the matrix model theory on a torus with a constant
antisymmetric tensor field as in \cite{K,H} looks like
the same as the one presented here. In spite of its formal
analogies as we have previously discussed, there are important
differences that affect not only to the spectral properties of the
theory but to the interpretation. Our bracket is not a Moyal
bracket but a non-constant symplectic one. At leading order in their
background expansion and for the case of a torus where the global analysis
is possible, both models coincide. In a more general
compactification manifold, i.e. a Riemanian surface, this is not
possible anymore. There exists a bundle of D2-D0 branes over the
manifold. The Darboux theorem ensures the existence of
 a finite covering of the manifold such that on each open set of the
 covering the supermembrane is equivalent to a $D2$ brane with constant
 $B$ field. These charts are glued by
the gauge group of diffeomorphisms preserving the area and can be
seen as intersecting D2 branes with different constant flux that
have been recombined into an stable D2-D0 brane over the
nontrivial topology. The stability is assured by the stability of
the nontrivially wrapped supermembrane.\\ 
In general, intersecting
supersymmetric Dp-branes can decay under certain topologies on a
recombined metastable Dp brane by giving mass to scalar
fields. In
our case this is exactly what happens in the sense that the
presence of central charge or in its dual version, the flux, gives
mass to the transverse scalar fields that parametrize the position
of the compact D2 brane in the target space. Nevertheless in this case the
condition of stability is much stronger that in other ones, since it is
assured by the stability of the supermembrane with central charge. The
recombination process has been used to break symmetries of gran
unified models in order to obtain more realistic ones, and they are
understood as a Higgs mechanism. Here the worldvolume gauge fields are
the responsible of this induced Higgs mechanism. As they are only
defined on the worldvolume they do not get in conflict with Poincare invariance.\\
There is another interesting interpretation that naturally emerge
from our picture. The supermembrane with central charge, can be
understood because of its spectral properties as a fundamental
object. We
speculate that it eventually could lead to reproduce part of the spectrum that
we observe. Phenomenological approach have been done in the context of
braneworlds. There, Dp-branes have  $p\ge 3$ to reproduce
Minkowski space where strings propagate. The spectrum is generated by the
strings whose  ends are attached in different Dp-branes as well as closed
string sector carrying gravity. Here the point of view is different, we
are thinking about the possibility of taking, the supermembrane with
central charge, as a fundamental object
propagating in the target space which is restricted in $4D$ to be
Minkowski.

In this direction \cite{D, S} studied wrapped $D2$ branes on
$\frac{C^{2}}{Z_{n}}\times X$ spaces. It
contains N 2-cycles and leaves $N^{2}$ possible wrapping configurations of
$D2$ branes. The wrapped $D2$ branes around singularities become
massless and are able to reproduce the gauge sector. In other type
of singularities it is also possible to obtain chiral matter. The gravity
sector is carried by closed strings.\footnote{We thank A. Uranga for
a clarifying discussion at this point}.
In the same spirit, we can consider several $M2$ nontrivially wrapped over
generic 2-cycles minimally inmersed in the target space that collapse
around adecuate singularities able to reproduce the matter spectrum. The
gravity sector is supossed to be carried by the supermembrane in 11D
propagating in the target space since the
supermultiplet of supergravity is conjectured to be the ground state
of the supermembrane. This sector would contain instabilities so 
a further study of the complete theory taking all of the sectors into
account would  be needed to have
good insight of the theory. We leave all of these
questions for a future work.

\section*{Acknowledgments}
We thanks A.Font, J. Bellor\'{\i}n, C. G\'omez, A. Gonz\'alez-Arroyo,
K. Lansteiner, E. L\'opez, A. Uranga, for enlighting discussions.
M.P.G.M. is supported by a postdoctoral grant of the Consejer\'{\i}a de
Educaci\'{o}n, Cultura y Deportes de la Comunidad Aut\'onoma de la
Rioja, (Spain). A.R. thanks the kind invitation to the IFT-UAM (Spain)
where part of this work was done.

\end{document}